\documentstyle[twoside,fleqn,espcrc2]{article}

\newcommand{\AmS}{{\protect\the\textfont2
  A\kern-.1667em\lower.5ex\hbox{M}\kern-.125emS}}

\title{Softly Broken SQCD: in the Continuum, on the Lattice, on the Brane }

\author{Nick Evans\address{Department of Physics, Boston University, 590
    Commonwealth Avenue, Boston, MA 02215, USA. E-mail:
    nevans@physics.bu.edu.}%
        \thanks{Much of this work was performed in collaboration with
         S.D.H. Hsu and M. Schwetz and was funded in part under DOE
         contract DE-FG02-91ER404676. The author would also like to
         thank R. Sundrum, M. Schmaltz, E. Poppitz, C. Johnson and 
         A. Shapere for
         useful discussions.}}

\begin{document}

\begin{abstract}
We review progress in studying the solutions of SQCD in the presence of
explicit, soft SUSY  breaking terms. Massive N=1 SQCD in the
presence of a small gaugino mass leads to a controlled solution which
has interesting phase structure with changing $\theta$ angle
reminiscent of the QCD chiral Lagrangian. Current attempts to test the
solutions of pure glue SQCD on the lattice require a theoretical
understanding of the theory with small gaugino mass in order to 
understand the  approach to the SUSY point. We provide such a
description making predictions for the gaugino condensate and lightest
bound state masses. Finally we briefly review recent D-brane
constructions of 4D gauge theories in string theory including a
non-supersymmetric configuration. We identify this configuration with 
softly broken N=2 SQCD.
\end{abstract}

\maketitle

\section{INTRODUCTION}

Our understanding of supersymmetric QCD (SQCD) has grown rapidly over
the last four years \cite{seiberg}. There are now 
solid examples of four dimensional gauge
theories that confine by the dual Meissner effect, exhibit chiral
symmetry breaking and even theories that give rise to massless
composites. One natural question to ask is how these phenomena relate to
non-supersymmetric theories and in particular to QCD. Some progress has
been made in including perturbing soft SUSY breaking interactions whilst
retaining the ``exactness'' of the supersymmetric results 
\cite{soft1,lattice,soft2,soft3}. We review
some of those continuum results and their importance to lattice
simulations of pure glue SQCD. Finally we discuss non-supersymmetric
``brane'' configurations in string theory and their field theory 
interpretation, identifying the corresponding soft SUSY breaking
operators.


\input epsf
\newwrite\ffile\global\newcount\figno \global\figno=1
\def\writedefs{\immediate\openout\lfile=labeldefs.tmp \def\writedef##1{%
\immediate\write\lfile{\string\def\string##1\rightbracket}}}
\def\writestoppt{}\def\writedef#1{}

\def\figin{\epsfcheck\figin}\def\figins{\epsfcheck\figins}
\def\epsfcheck{\ifx\epsfbox\UnDeFiNeD
\message{(NO epsf.tex, FIGURES WILL BE IGNORED)}
\gdef\figin##1{\vskip2in}\gdef\figins##1{\hskip.5in}
\else\message{(FIGURES WILL BE INCLUDED)}%
\gdef\figin##1{##1}\gdef\figins##1{##1}\fi}

\def\figinsert{}
\def\ifig#1#2#3{\xdef#1{}
\writedef{#1\leftbracket fig.\noexpand~\the\figno}%
\figinsert\figin{\centerline{#3}}\medskip\centerline{\vbox{\baselineskip12pt
\advance\hsize by -1truein\center\footnotesize{  } #2}}
\bigskip\endinsert\global\advance\figno by1}
\def\footnotefont{}\def\endinsert{}

\section{SOFT BREAKINGS IN N=1 SQCD}

We begin from the N=1 $SU(N_c)$ SQCD theories with $N_f$ flavors
described by the UV Lagrangian
\begin{eqnarray}
{\cal L}& =& K (Q^\dagger_i Q_i + \tilde{Q}^\dagger_i
\tilde{Q}_i)|_D + {1 \over 8 \pi}
Im \tau W^\alpha W^\alpha|_F \nonumber \\
&&+ 2 Re\, m_{ij} Q_i \tilde{Q}_j|_F 
\end{eqnarray}
where $Q$ and $\tilde{Q}$ are the standard chiral matter superfields and
$W^\alpha$ the gauge superfield. The coupling $K$
determines the kinetic normalization of the matter fields. The
gauge coupling $\tau = \theta/2 \pi + i 4 \pi/g^2$  defines
a dynamical scale of SQCD:
$\Lambda^{b_0} = \mu^{b_0} exp( 2\pi i \tau)$,
with $b_0 = 3 N_c - N_f$ the one loop coefficient of the SQCD
$\beta$-function.
And, finally, $m$ is  a supersymmetric mass term for
the matter fields. We may raise these couplings to the status of spurion
chiral superfields which are then frozen with scalar
component vevs. 

Soft supersymmetry breaking parameters may be introduced through
the F-component of the spurion coupling fields. A gaugino mass may be
generated from the gauge coupling $\tau$, $F_\tau = i
8 \pi m_\lambda$

\begin{equation} 
{1 \over 8 \pi} Im \tau WW|_F \rightarrow Re m_\lambda \lambda \lambda
\end{equation}

Scalar masses and interactions may be introduced through the mass
spurion and kinetic normalization $K$. $F_m \neq 0$ gives

\begin{equation} 
Re m \tilde{Q} Q|_F \rightarrow Re F_m A_{\tilde{Q}} A_Q
\end{equation}
and allowing a component of $K = m_Q^2 \theta^2 \bar{\theta}^2$
\begin{equation} 
K Q^\dagger e^V Q|_D \rightarrow m_Q^2 |A_Q|^2
\end{equation}

It is particularly useful to write the soft breakings as the components
of the spurion fields because the symmetries of the SQCD model are left
unaltered even in the softly broken model. 

The SQCD theory
without a mass term has the symmetries
\begin{equation}
\begin{tabular}{ccccc}
&$SU(N_f)$ & $SU(N_f)$ & $U(1)_B$ & $U(1)_R$\\
$Q$ & $N_f$ & 1 & 1 & ${N_f - N_c \over N_f}$\\
$\tilde{Q}$ &1& $\bar{N}_f$ & -1 & ${N_f - N_c \over N_f}$\\
$W^\alpha$ & 1 & 1 & 0 & 1\end{tabular}
\end{equation}
The mass term breaks the chiral symmetries to the vector symmetry. The
classical $U(1)_A$ symmetry on the matter fields is anomalous and,
if there is a massless quark, may be used to rotate away the $\theta$
angle. In the massive theory the flavor symmetries may be used to
rotate $m_{ij}$ to diagonal form and the anomalous $U(1)_R$ symmetry
under which the $Q$s have charge $+1$ may be used to rotate $\theta$ on
to the massless gaugino. Including
the spurion fields the non-anomalous $U(1)_R$ symmetry charges are
\begin{equation}\label{sym}
\begin{tabular}{ccccc}
$W$ & $Q$ & $\tilde{Q}$ & $\tau$ & $m$ \\
1 & ${N_f - N_c \over N_f}$ & ${N_f - N_c \over N_f}$ & 0 & ${2N_c \over
  N_f}$ \end{tabular}
\end{equation}
The anomalous symmetries may be restored to the status of symmetries of
the model if we also allow the spurions to transform. The appropriate
charges are
\begin{equation}
\begin{tabular}{cccccc}
&$W$ & $Q$ & $\tilde{Q}$ & $\Lambda^{b_0}$ & $m$\\
$U(1)_R$ & 1 & 0  & $ 0 $ & $2(N_c-N_f)$ &  2  \\
$U(1)_A$ & 0 & 1  & 1       & $2N_f$             & -2              
\end{tabular}
\end{equation}
The $m_{ij}$ spurions also transform under the
chiral flavor group.

These symmetries and supersymmetry remain symmetries of the model no
matter which components of the spurions are non-zero and hence they may
be used to determine the low energy theory of the softly broken
models (there is an assumption that there is not a phase transition to 
a totally different set of variables as soon as supersymmetry is
broken). 
The use of these symmetries is completely analgous to the use of
chiral symmetry in QCD to find the mass dependence of the QCD chiral
Lagrangian. 

For pure supersymmetric models the potential minima may be found from
the superpotential alone which is holomorphic in the fields and
spurions. The exact results for the far IR behaviour of the theories
result from the very limited number of possible terms compatible with
the symmetries. There is an immediate problem in the softly broken
theories though which is that 
scalar masses may be generated from non-holomorphic
Kahler terms. For example

\begin{equation}
\tau^\dagger \tau Q^\dagger Q|_D \rightarrow |F_\tau|^2 |A_Q|^2
\end{equation}

Thus for example if one begins from an SQCD theory with a moduli space
in the scalar vevs the minima of the potential in the softly broken
model will depend on these unknown terms. In particular one does not
know the sign of these mass terms; a negative mass would indicate a
higgs mechanism and a mass gap, a positive mass would leave the scalar's
fermionic partner massless.

A solution to this problem \cite{soft1} 
is to begin with an SQCD theory in which the
scalars have supersymmetric masses. For small soft breakings relative to
that mass the unknown Kahler terms may be neglected. As the simplest
example consider SQCD with a mass term for the matter fields. The 
resulting theories
have a mass gap on the scale $m$ and the induced meson $M_{ij}= Q^i
\tilde{Q}_j$ vev is determined independently of $N_f$ by holomorphy
\begin{equation}\label{Slimit}
M_{ij} = \Lambda^{{3N_c - N_f \over N_c}}  (detm)^{1/N_c}\left( {1 \over
  m} \right) _{ij} = |M_{ij}| e^{i\alpha}~~~
\end{equation}
The resulting supersymmetric theories have $N_c$ distinct vacua
corresponding to the $N_c$th roots of unity, $\alpha = 2n\pi/N_c$
(as predicted by the Witten
index). Note that for the theories
with magnetic duals putting masses in for all flavors breaks the dual
gauge group completely. For simplicity henceforth we shall take $m_{ij}$
to be proportional to the identity matrix; in this basis $\langle
M_{ij} \rangle$ is also proportional to the identity matrix.

These massive theories may be softly broken in a controlled fashion.
If the spurion generating the soft breaking enters
the superpotential linearly then we may obtain desirable results when that
spurion's F-component $F \ll m \ll \Lambda$. Any Kahler term contributions to
the scalar potential take the form $F_X^\dagger F_Y$ with $X$ and $Y$
standing for generic fields or spurions. In the supersymmetric limit all
F-components are zero and will grow as the vacuum expectation value of the
soft breaking spurion.
These Kahler terms are therefore higher order in
the soft breaking parameter than the linear term from the
superpotential. The unknown corrections to the squark masses in the
theory are subleading to the masses generated by the supersymmetric mass
term and hence we may determine the potential minima at lowest order.

As an example we introduce a gaugino mass through the spurion $\tau$.
In the IR theory $\tau$ enters through the strong interaction scale
$\Lambda$ which occurs linearly in the superpotential of the
theory. Taking $F_{\tau} \ll m \ll \Lambda$ we may determine
the vacuum structure. The IR superpotential terms compatible with the
symmetries of the theory involving $\Lambda$ are
\begin{equation}
Re[ m M_{ij} + ({\rm det} M_{ij})^ { 1 \over (N_f - N_c) } \Lambda^{(3N_c-N_f)
  \over (N_c-N_f)}]
\end{equation}
where the final term results from non-perturbative effects in the broken gauge
group. At lowest order in perturbation theory the vev of $M_{ij}$ is
given by (\ref{Slimit}) which also contains $\Lambda$ and hence has a
non-zero F-component. Including $F_\tau$ and performing the superspace
integral we obtain up to a coefficient the following corrections
to the potential that
break the degeneracy between the $N_c$ SQCD vacua
\begin{eqnarray}
\label{gpot}
\Delta V & = & Re\left[ m^{N_f/N_c} 8 \pi m_\lambda \Lambda^{(3N_c-N_f)/
                N_c}\right]\\
& = \nonumber & \left|m^{N_f/N_c} 8 \pi m_\lambda \Lambda^{(3N_c-N_f)\over
                N_c}\right| \\
&& \left. \right. \hspace{0.7cm} \times 
\cos[ ~ {\theta_{phys} \over N_c} ~+~ \alpha ~]
\end{eqnarray}
where $\alpha$ are the $N_c$th roots of unity and
$\theta_{phys}$ is the physical $\theta$ angle in which the
physics must be $2 \pi$ periodic
\begin{equation}
\theta_{phys} ~=~ \theta_0 ~+~ N_c ( \theta_{m_\lambda}+  N_f \theta_m )
\end{equation}
The gaugino mass has explicitly broken the anomalous U(1) symmetries of
the SQCD model and hence the $\theta$ angle may not be rotated away.
There is also an additional contribution to the vacuum energy
arising from the gaugino condensate. Using the Konishi
anomaly \cite{KA}, we see that it has the same form as
(\ref{gpot}).

The supersymmetry breaking contributions to the potential 
break the degeneracy
between the $N_c$ supersymmetric vacua. The model has interesting phase
structure as the bare $\theta$ angle is changed. There are phase transitions
as $\theta_{phys}$ is varied,  occurring
at $\theta_{phys} ~=~ $(odd)$\pi$.
This behavior can be compared
with the $\theta$ angle dependence of the QCD chiral Lagrangian \cite{chiral}
for which there are $N_f$ distinct vacua which interchange through first
order phase transitions at $\theta =$(odd)$\pi$.

Unfortunately if we wish to keep control of the low energy solution we
are forced to keep the soft breakings small and we can not decouple the
superpartners to recover non-supersymmetric QCD. There is however
one conclusion for QCD that we can tentatively draw from this
analysis. In these models the form of the confined effective
theory changes smoothly with the $\theta$ angle and there is no sign of a
break down of confinement as suggested in \cite{schierholz}. This lends some
support to the assumption \cite{chiral} that the chiral Lagrangian remains
the correct discription of QCD in the IR even at non-zero $\theta$.

\section{PURE GLUE SQCD AND LATTICE TESTS}

Although the techniques for solving the supersymmetric and softly broken
theories described above provide an extremely plausible
picture of the low-energy dynamics of these models, one may feel a 
little discomfort at the absence of direct 
non-perturbative tests of the results.
An obvious possibility is that these models could be simulated directly 
on the lattice. Some initial work in these directions has already been
performed in \cite{Montvay}.Unfortunately, as is well known, 
lattice regularization violates 
supersymmetry \cite{CV}, and a special fine-tuning is required to
recover the SUSY limit (this is analogous to the case of chiral 
symmetry in lattice QCD). Away from the SUSY point, the continuum
limit of the lattice theory is described by a model with explicit SUSY  
violating
interactions. In some cases, these violations may correspond only
to soft breakings, 
although this is not guaranteed in general. 

Pure glue SQCD is a simple theory with only one parameter,
the gauge coupling. The only low-dimension (renormalizable) 
SUSY violation allowed by gauge invariance is a gaugino mass,
which is a soft violation. Therefore, the continuum limit of the
lattice regularized version of SYM is simply SYM with a massive gaugino.
The SUSY limit can be reached by fine-tuning the lattice parameter
corresponding to a bare gaugino mass. 
In order to understand this limit as well as possible, we will study continuum
SYM with explicit gaugino mass \cite{lattice}, and derive some relations describing
the approach to the SUSY limit. Several non-trivial predictions can
be made regarding the vacuum energy and of the behavior of the 
gaugino condensate. A less rigorously derived description of the 
lightest bound states of SYM theory has also been proposed in the 
literature \cite{VY} from which predictions for the masses 
of the gluino-gluino and glue-gluino bound states and their splittings
away from the supersymmetric point may be obtained. 

The bare Lagrangian of  SYM  with $SU(N_c)$ gauge group  
is
\begin{equation}
\label{SYM}
{\cal L} ~=~ \frac{1}{g_0^2}\left[ \,-\frac{1}{4} 
G_{\mu\nu}^a G_{\mu\nu}^a
~+~ i\lambda_{\dot\alpha}^\dagger D^{\dot\alpha\beta}\lambda_\beta
\right] ~. 
\end{equation}
This model possesses a discrete global $Z_{2N_c}$ symmetry, a 
residual non-anomalous subgroup of the anomalous chiral $U(1)$. The
theory is believed to generate a gaugino condensate and have a mass
gap.

In supersymmetric notation the Lagrangian (\ref{SYM}) can be written as
\begin{equation}
\label{SSYM}
{\cal L} ~=~ \int d^2 \theta~
\frac{1}{8\pi} Im\, \tau_0 W^{\alpha}W_{\alpha}~, 
\end{equation}
where the gauge coupling is defined to be
$\tau_0 = \frac{4\pi i}{g^2_0} + \frac{\theta}{2\pi}$. 
Note that $\Lambda^{b_0} = e^{2\pi i \tau_0} \mu^{b_0}~$ is 
explicitly $2\pi$-periodic in the
$\theta$-angle.

To derive the low energy effective theory of SQCD we note that there are
two anomalous symmetries of the theory, $U(1)_R$ and scale
invariance. In
fact their anomalies are related since their currents belong to the same
supermultiplet. These symmetries can be restored in an enlarged theory
provided we allow the spurion couplings to transform:
\begin{eqnarray}
U(1)_R: \hspace {1cm} &&
W(x,\theta) \rightarrow e^{i\zeta} W(x, \theta e^{i\zeta}) \nonumber \\
&& \Lambda \rightarrow \Lambda e^{i2 \zeta/3} \nonumber 
\end{eqnarray}
\begin{eqnarray}
{\rm Scale} \hspace{0.1cm} {\rm  Invariance}: 
&& W(x,\theta) \rightarrow e^{3\xi/2} W(xe^\xi, \theta e^{\xi/2})
\nonumber \\
&& \Lambda \rightarrow \Lambda e^\xi ~~~~~.\nonumber
\end{eqnarray}

We may now determine the general form of the partition function 
(assuming a mass gap) as a
function of $\tau$ subject to these symmetries. The only possible terms
are
\begin{equation}
\label{SZ}
Z[\tau] = {\rm} \hspace{0.1cm} {\rm exp} ~ iV \left[  {9 \over \alpha}  
\Lambda^\dagger \Lambda|_D + ( \beta
\Lambda^3|_F +h.c.) \right]
\end{equation}

The numerical coefficients $\alpha$ and $\beta$ remain undetermined from
the above symmetry arguments. $\beta$ may be determined from the results
for SQCD with massive quarks where for $N_f = N_c -1$ the full gauge
symmetry may be higgsed and the coefficient of the superpotential term
calculated by  perturbative instanton methods. 
We find $\beta = N_c$ \cite{cordes}.

These strong arguments lead to two predictions for the
condensates of the SYM theory. The source $J$ for the
gaugino correlator $\lambda \lambda$ occurs in the same position as the 
F-component of $\tau$ and is hence known. 
There are two independent correlators
\begin{eqnarray}
\label{cond}
\langle \lambda \lambda \rangle  & = & - 32 \pi^2 \Lambda^3\nonumber \\
\langle \bar{\lambda} \bar{\lambda} \lambda \lambda\rangle &  = & 
{- 1024 i \pi^4 \over \alpha N_c^2} |\Lambda|^2 / V~~~.
\end{eqnarray}

The IR theory has a gaugino condensate $\simeq \Lambda^3$, with phase
$2\pi i\tau /N_c$ and hence there are $N_c$ degenerate 
vacua associated
with the $N_c$th roots of unity. Below, therefore, $\Lambda^3$ 
is an $N_c$ valued constant with phases $n 2\pi i /N_c$ where $n$ runs from
$0...$ $N_c-1$.

\subsection{Soft Supersymmetry Breaking}

We may induce a  bare gaugino mass through a non zero
F-component of the bare gauge coupling 
$\tau = \tau_0 + 8 \pi i m_\lambda \theta \theta$ 

In the IR theory $\tau$ enters through the spurion
$\Lambda$ which occurs linearly in the superpotential of the
theory. Thus there will be a correction to the potential
of the form:
\begin{equation}
\label{correction}
\Delta V ~=~32 \pi^2 Re ( m_\lambda  \Lambda^3)  - {256 \pi^4 
  \over \alpha N_c^2} |m_\lambda \Lambda|^2
\end{equation}
Terms with superderivatives acting on the spurion field
can also give rise to contributions to the potential but these are
higher order in an expansion in $m_\lambda/\Lambda$. The shift in
the potential energy of the $N_c$ degenerate vacua of the SYM theory
at linear order in $m_\lambda$ is known and we may determine the vacuum
structure 
\begin{equation}
\label{deltaV}
\Delta V = 32 \pi^2 |m_\lambda \Lambda^3|~ \cos \left[ {2 \pi n \over N_c} +
\theta_{m_\lambda} \right]
\end{equation}
For small soft 
breakings, $m_\lambda \ll \Lambda$, where the linear term dominates, 
the degeneracy between the SYM vacua is broken favoring one vacuum 
dependent on
the phase of the gaugino mass.
The coefficient in the energy shift
is a test of the exact superpotential in (\ref{SZ}).

We may also determine the leading shift in the gaugino condensate
\begin{equation}
\label{Stau}
\langle \lambda \lambda \rangle 
~=~ - 32 \pi^2 \Lambda^3 \, 
~+~ \frac{512 \pi^4}{\alpha N^2_c} m_\lambda^* |\Lambda|^2~,
\end{equation}
which depends on the unknown parameter $\alpha$. Strictly speaking
there are also divergent contributions to this quantity which are
proportional to $m_\lambda$ times the cut-off squared. 

Reinserting the bare $\theta_0$ angle into the expression for the shift in
vacuum energy we find 
\begin{equation}
\Delta V = 32 \pi^2 |m_\lambda \Lambda^3|~ \cos \left[ {2 \pi n \over N_c} +
\theta_{m_\lambda} + {\theta_0 \over N_c} \right]
\end{equation}
As $\theta_0$ is changed first order phase transitions occur at
$\theta_0 = ({\rm odd}) \pi$ where two of the $N_c$ SYM vacua
interchange as the minimum of the softly broken theory.

\subsection{The Lightest Bound States}

An alternative description of the low energy behaviour of SYM theory has
been presented by Veneziano and Yankelowicz \cite{VY} which attempts to
describe the lightest bound states of the theory. The form of their
effective action can be rigourously obtained from the discussion above.
Since the source $J$ for $WW$ occurs in the same places as the
coupling $\tau$ we also know the source dependence of $Z$. If we wish we
may Legendre transform $Z[\tau,J]$ to obtain the effective potential for
the classical field
\begin{equation}
S \equiv 
-\frac{1}{32\pi^2}\,\mbox{Tr}\,\langle W^2 \rangle~~.
\end{equation}
We find 
\begin{eqnarray}
\label{VY}
\Gamma[\tau,S] & = & {9 \over \alpha} \left( \bar S S
\right)^{1/3}\Big|_D\\
&&  ~+~ 
N_c \left( S - S\ln (S/ \Lambda^3) \right)\Big|_F+ \, \mbox{h.c.}~ \nonumber
\end{eqnarray}
So derived this effective action contains no more information than
(\ref{SZ}) simply being a classical potential whose minimum determines
the vev of $S$ and we find, by construction, Eq(\ref{cond}).

A stronger interpretation can also though be given to the VY action. If
we assert that the lightest bound states of the theory are those that
interpolate in the perturbative regime to the field $WW$, and hence share
the same symmetry properties, then those symmetries again reproduce the
VY action for those lightest fields. To obtain the physical states
one performs 
an appropriate rescaling of the $S$-field 
\begin{equation}
\label{Sres}
\Phi~=~ \frac{3}{\sqrt{\alpha}} \,S^{1/3}
\end{equation}
in the Lagrangian (\ref{VY}) 
to make the kinetic term canonical 

\begin{eqnarray}
\label{VYLR}
{\cal L} &~=~& \left( \bar \Phi \Phi \right)\Big|_D~+~
\frac{a^{3/2}N_c}{9}\left(\frac{1}{3}\Phi^3 \right.
\nonumber \\
&& \left. -  \Phi^3 
\ln{(\frac{a^{1/2}}{3}\frac{\Phi}{\Lambda} )}\right)\Big|_F
~+~ \, 
\mbox{h.c.}~
\end{eqnarray}

In fact, as pointed out and corrected in \cite{Shifman}, this effective
Lagrangian is not complete since it does not possess the full $\rm Z_{2N_c}$
symmetry of the quantum theory. To restore that symmetry the extra term
\begin{equation}
\Delta {\cal L} = { 2 \pi i m \over 3} \left( S - \bar{S} \right)
\end{equation}
where $m$ is an integer valued Lagrange multiplier must be added. For
the $n=0$ vacuum with vanishing phase this extra term vanishes
and the VY model above is recovered. We shall concentrate on that vacuum.

One must worry about possible mixing between $\phi$ and the next
massive state with the same quantum numbers but it seems reasonable that
this state may be significantly heavier and hence may be neglected. We
shall move on and use the VY action as a description of the lightest
states to make predictions about the masses of those states. A
lattice simulation will hopefully test these predictions and shed light
on whether the action is indeed the correct description.

The straightforward evaluation of bosonic ($\lambda \lambda $) 
and fermionic ($g\lambda$) excitation masses  
around
the minimum from Eq(\ref{VYLR}) gives
\begin{equation}
\label{susymass}
m_{\lambda \lambda}~=~m_{g\lambda}~=~N_c \alpha \Lambda~.
\end{equation}

It is important to stress that these masses are not the
physical masses of the bound states. Rather, they are zero-momentum
quantities, which are related to the physical ones by wave function
factors $Z(p^2 = m_{phys}^2)$. These wave function factors result from
higher-derivative Kahler terms in ${\cal L}$, and are unknown.

A soft breaking gaugino mass may again be introduced through the
F-component of the spurion $\Lambda$. 
We can calculate the shifts in the masses of the bound
states. The two scalar fields and the fermionic field are split in
mass
\begin{eqnarray}
M_{\rm fermion} & = & N_c \alpha \Lambda - {16 \pi^2 m_\lambda 
\over  N_c}  \nonumber \\
M_{\rm scalar} & = \nonumber & N_c \alpha \Lambda - {56 \pi^2 m_\lambda 
\over 3 N_c}\\
M_{\rm p-scalar} & = & N_c \alpha \Lambda - {40 \pi^2 m_\lambda  
\over 3 N_c}
\end{eqnarray}

The physical masses are again related to these quantities by
unknown wave function renormalizations $Z$ which arise from  Kahler terms, 
$$
M_{\rm physical} 
~ \equiv ~ Z ~ M ~~. 
$$
Fortunately, we know that
in the SUSY limit the wavefunction factors are common
within a given multiplet. This degeneracy holds even after the vev of
the field is shifted by the soft breakings since a shift in the vev
alone (without SUSY breaking) leaves the physical masses degenerate
within a multiplet.
We also know that the {\it relative change} in these Kahler terms is  
${\cal O} (f_\tau^2)$, 
and hence can be ignored at leading order in
the soft breakings. Therefore, we may 
still obtain a prediction for the rate of change of
the ratios of the physical masses, 
\begin{eqnarray}
\label{bm}
\bar{M} (m_\lambda) & ~\equiv ~ & { Z (m_\lambda) M(m_\lambda )- Z(0) M(0)  
\over Z(0) M(0) }~~,
\nonumber\\
& \simeq & {\partial M \over \partial m_\lambda} 
\left[ \frac{1}{M}   + 
\frac{1}{Z}  {\partial Z \over \partial M } \right]  m_\lambda 
\end{eqnarray}
near the SUSY limit. 
The factor in brackets is common within a given multiplet.
Since the quantity $Z(m_0)$ is unknown, we can only predict
the {\it ratios} of $\bar{M}$ at the SUSY point or equivalently the
ratios
of $\partial M / \partial m_\lambda$.

Finally we note, as pointed out in \cite{Shifman},
that the VY model apparently has an extra SUSY vacuum corresponding to
$\langle \phi \rangle = 0$. At this point $\langle S \rangle$
is singular and so it is not clear how to interpret this vacuum. Shifman
and Kovner have proposed that the vacuum is real and represents some
conformal, $Z_{2N_c}$ preserving 
point of the theory. It would be interesting to look for this
vacuum in lattice simulations but unfortunately as can be seen from
(22)  there is no value of soft breaking mass for which such a
vacuum would be the global minimum. This will make it difficult to
observe in lattice simulations.

\section{FIELD THEORY DUALITY AND SOFT BREAKING FROM STRING THEORY}

The most recent progress in understanding SQCD has come from string
theory. In type IIA string theory D-brane constructions can be made
that realize 4D field theories in the world volume of one of the
D-branes \cite{elitzur,witten,brand}. 
The standard construction is to suspend $N_c$ D4-branes
between two NS5-branes to generate an $SU(N_c)$ gauge symmetry in the
D4-branes' world volume. $N_f$ D6-branes intersecting the D4-branes contribute
vector matter multiplets transforming under the gauge symmetry and a
gauged flavor symmetry $SU(N_f)$. For example an N=2 configuration may
be described as follows \cite{witten}.

$\left. \right.$ \hspace{-0.15cm}
\begin{tabular}{|c|c|c|c|c|c|c|c|c|}
\hline
 & $\#$ & $R^4$ & $x^4$ & $x^5$ &  $x^6$ & $x^7$ & $x^8$ & $x^9$ \\
\hline 
NS & 2 & $-$ & $-$ & $-$  &  $\bullet$ & $\bullet$ & $\bullet$ & $\bullet$ \\
\hline
D4 & $N_c$ & $-$  & $\bullet$ &  $\bullet$ &  $[-]$ & $\bullet$ & 
                              $\bullet$ & $\bullet$ \\
\hline
D6 & $N_f$ & $-$  & $\bullet$ &  $\bullet$ &  $\bullet$  & $-$ & 
                              $-$ & $-$ \\
\hline
\end{tabular} \vspace{0.3cm}

$R^4$ is the space $x^0-x^3$ which will correspond to the 4D space in
which the $SU(N_c)$ gauge theory will live. A dash $-$ represents a
direction along a brane's world wolume while a dot $\bullet$ is
transverse. For the special case of the D4-branes' $x^6$ direction,
where a world volume is a finite interval, we use the symbol $[-]$. On
scales much greater than the $L_6$ distance between the NS5s the fourth
space like direction of the D4-branes generates  the coupling of the
gauge theory in an effective 3+1D theory. 

A rotation of the two NS5s relative to each other breaks the
supersymmetry of the configuration further. An N=1 SQCD theory results
from the configuration.

$\left. \right.$ \hspace{-0.15cm}
\begin{tabular}{|c|c|c|c|c|c|c|c|c|}
\hline
 & $\#$ & $R^4$ & $x^4$ & $x^5$ &  $x^6$ & $x^7$ & $x^8$ & $x^9$ \\
\hline 
NS & 1 & $-$ & $-$ & $-$  &  $\bullet$ & $\bullet$ & $\bullet$ & $\bullet$ \\
\hline
NS & 1 & $-$  & $\bullet$ & $\bullet$  &  $\bullet$ & $\bullet$ & $-$ & $-$ \\
\hline
D4 & $N_c$ & $-$  & $\bullet$ &  $\bullet$ &  $[-]$ & $\bullet$ & 
                              $\bullet$ & $\bullet$ \\
\hline
D6 & $N_f$ & $-$  & $\bullet$ &  $\bullet$ &  $\bullet$  & $-$ & 
                              $-$ & $-$ \\
\hline
\end{tabular} \vspace{0.3cm}

This configuration first considered in \cite{elitzur} was used to derive
the field theory duality for $N_f > N_c$. It can be drawn pictorially as

$\left. \right.$  \hspace{-0.4in}\ifig\prtbdiag{}
{\epsfxsize 7truecm\epsfbox{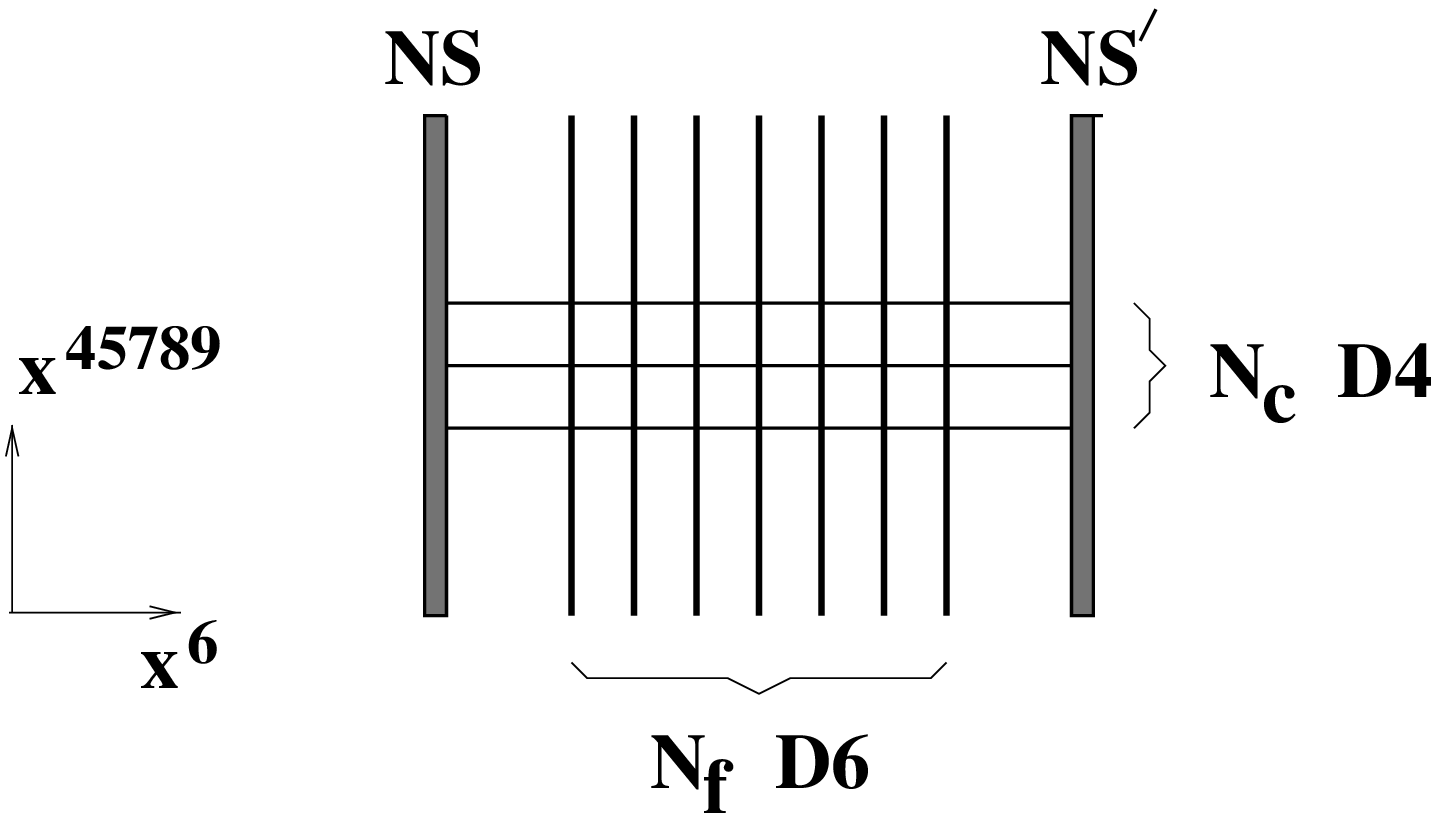}} \vspace{-1cm}

The duality was realized by
the motion of the two NS5s through each other in the $x^6$
direction. This motion corresponds to changing the coupling constant of
the classical gauge theory. In the quantum theory where there is an IR
fixed point such a change of the UV coupling leaves the IR physics
invariant and hence it is claimed that the configurations after these
 motions continue to
describe the same quantum theory. Everytime a NS5 passes through a
D6-brane an extra D4-brane must be extruded between them in order to
preserve the number of matter fields in the theory on the $N_c$
D4-branes. When the two NS5s pass through each other the string theory
and the field theory pass through a strong coupling regime. There is
again a conservation rule for the number of D4-branes connecting the two
NS5s. The resulting motion corresponds to the final configuration

$\left. \right.$  \hspace{-0.4in}\ifig\prtbdiag{}
{\epsfxsize 7truecm\epsfbox{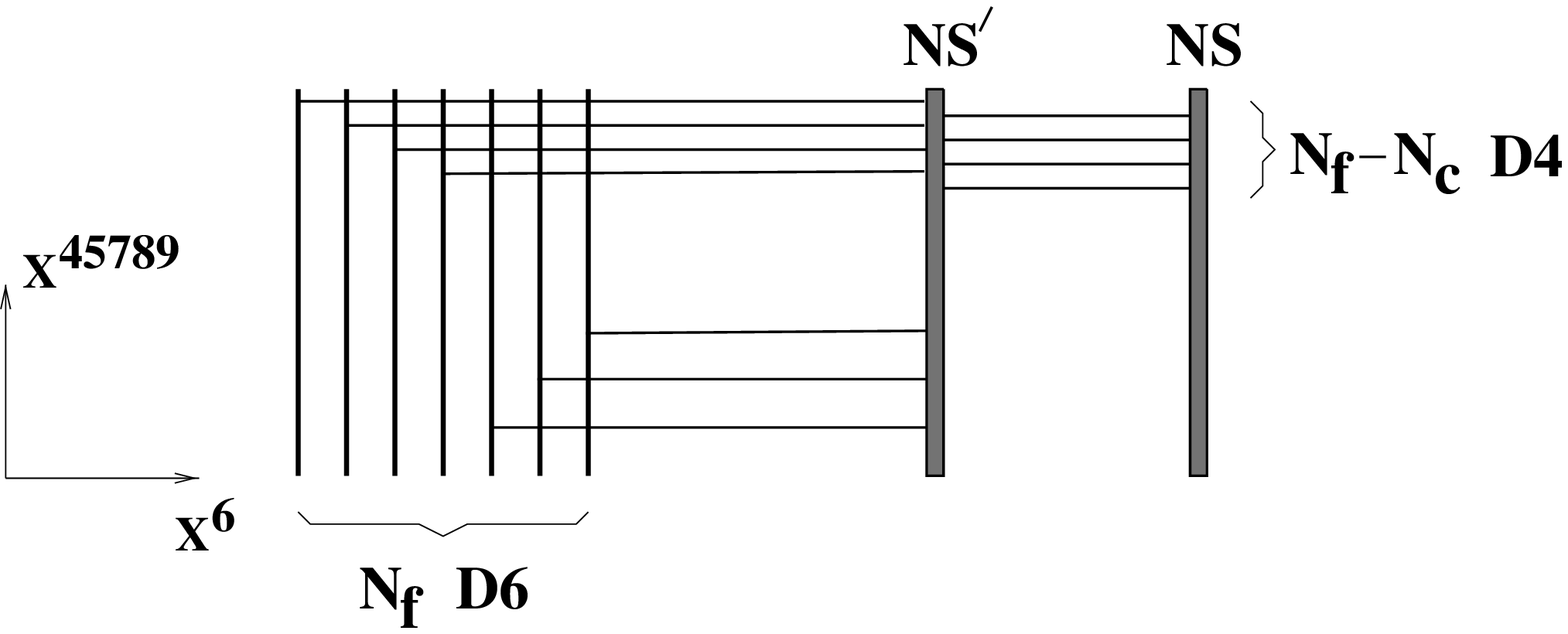}} \vspace{-1cm}

This configuration has an $SU(N_f-N_c)$ gauge symmetry, $N_f$ quark
fields and $N_f^2$ ``mesons'' associated with the freedom of motion of
the connections between the $N_f$ D4 branes to the left and the
D6-branes in the $x^8$, $x^9$ directions.

More recently \cite{witten} it has been realized that for the theories without
dualities the dynamically generated superpotential of the theories may
be derived from the curves describing the brane configurations when they
are extended to include the M-theory compactified dimension. 

Our interest here is in the possibility that these brane configurations
may be able to shed light on non-supersymmetric configurations. As
pointed out in \cite{brand} a different rotation of the N=2 configuration
leads to an N=0 configuration. We will only consider the pure glue case
in the hope of identifying the resulting field theories.

$\left. \right.$ \hspace{-0.15cm}
\begin{tabular}{|c|c|c|c|c|c|c|c|c|}
\hline
 & $\#$ & $R^4$ & $x^4$ & $x^5$ &  $x^6$ & $x^7$ & $x^8$ & $x^9$ \\
\hline 
NS & 1 & $-$ & $-$ & $-$  &  $\bullet$ & $\bullet$ & $\bullet$ & $\bullet$ \\
\hline
NS & 1 & $-$  & $-$ & $\bullet$  &  $\bullet$ & $\bullet$ &
$\bullet$   & $-$ \\
\hline
D4 & $N_c$ & $-$  & $\bullet$ &  $\bullet$ &  $[-]$ & $\bullet$ & 
                              $\bullet$ & $\bullet$ \\
\hline
\end{tabular}\vspace{0.3cm}

This configuration describes an $SU(N_c)$ gauge theory with a real
adjoint scalar field corresponding to the freedom to separate the D4-branes
in the $x^4$ direction. Can we identify the SUSY breaking terms
introduced into the N=2 theory that leaves this N=0 theory? We expect
the N=2 SUSY in 10D string theory when broken by the string dynamics to
appear as spontaneous SUSY breaking in the low energy field theory
description. We also expect non-renormalizable operators to be
suppressed by the planck scale. Thus any SUSY breakings will be precisely
those of the form of soft breakings that may be introduced through the
vevs of spurion fields in the theory. The N=2 theory has a single
spurion field $\tau$ which is a member of a full N=2 spurion
multiplet. That multiplet has three real auxilliary fields that may
acquire SUSY breaking vevs, the complex F-component of the matter field
and the real D component of the gauge field. These
spurion breakings have been investigated in \cite{soft2} and give rise to the
bare UV soft breakings

\begin{eqnarray}
&{1 \over 8 \pi} Im ( F^* \psi_A^\alpha \psi_A^\alpha + F
\lambda^\alpha \lambda^\alpha + D  \psi_A^\alpha\lambda^\alpha ) \\
& - {|F|^2 + D \over 4 \pi Im \tau} (Im
a^\alpha)^2 & \nonumber 
\end{eqnarray}

These breakings indeed leave a massless real adjoint scalar in the
theory. We identify the three components of the spurion field with
rotations into the $x^7$, $x^8$ and $x^9$ directions. 

These softly broken theories have been studied \cite{soft2} 
for small soft breakings
in SU(2) gauge theory with up to three flavors and the theories have been
seen to confine and have a mass gap on the scale of the soft breaking
with a dual weakly coupled IR description. 
The N=0 theories with
matter fields appear to have the motions described above that move the
theory to a dual description. If the identification of the N=0 theories
is correct though there are no massless adjoint fermions and if  the
electric and dual theories had massless fermions would not match
anomalies. The duality presumably continues to hold in the spirit of
\cite{soft2} as a dual Meissner description of confinement.
Although the
holomorphic properties of supersymmetry are lost in these brane
configurations it remains to be seen whether they can shed further light
on softly broken theories.

\end{document}